\documentclass[english,journal]{IEEEtran}
\usepackage[T1]{fontenc}
\usepackage[latin9]{inputenc}
\usepackage{units}
\usepackage{amsmath}
\usepackage{amssymb}
\usepackage{graphicx}

\makeatletter
\newenvironment{lyxlist}[1]
	{\begin{list}{}
		{\settowidth{\labelwidth}{#1}
		 \setlength{\leftmargin}{\labelwidth}
		 \addtolength{\leftmargin}{\labelsep}
		 }}
	{\end{list}}

\@ifundefined{date}{}{\date{}}
\usepackage{cite}
\usepackage[printonlyused]{acronym}
\acrodef{AWGN}{additive white Gaussian noise}
\acrodef{ASE}{amplified spontaneous emission}
\acrodef{QAM}{quadrature amplitude modulation}
\acrodef{PAM}{pulse amplitude modulation}
\acrodef{SE}{spectral efficiency}
\acrodef{SNR}{signal to noise ratio}
\acrodef{TX}{transmitter}
\acrodef{RX}{receiver}
\acrodef{BER}{bit error rate}
\acrodef{SER}{symbol error rate}
\acrodef{NFT}{nonlinear Fourier transform}
\acrodef{BNFT}{backward NFT}
\acrodef{FNFT}{forward NFT}
\acrodef{I-FNFT}{incremental FNFT}
\acrodef{DF-FNFT}{decision-feedback FNFT}
\acrodef{DF-BNFT}{decision-feedback BNFT}
\acrodef{NFDM}{nonlinear frequency-division multiplexing}
\acrodef{OFDM}{orthogonal frequency-division multiplexing}
\acrodef{NIS}{nonlinear inverse synthesis}
\acrodef{DAC}{digital-to-analog converter}
\acrodef{ADC}{analog-to-digital converter}
\acrodef{GVD}{group velocity dispersion}
\acrodef{SMF}{single mode fiber}
\acrodef{NCG}{Nystrom conjugate gradient}
\acrodef{B2B}{back-to-back}
\acrodef{EDC}{electronic dispersion compensation}
\acrodef{MAP}{maximum a posteriori probability}

\usepackage[linesnumbered,ruled,vlined]{algorithm2e}
\usepackage{algorithmic}

\makeatother

\usepackage{babel}
\begin{document}
\title{Nonlinearity Compensation for Coherent Optical Satellite Communications}
\author{Stella Civelli\thanks{S.~Civelli is with the Cnr-Istituto di Elettronica e di Ingegneria
dell\textquoteright Informazione e delle Telecomunicazioni (CNR-IEIIT),
Pisa, Italy. S.~Civelli, M.~Secondini and E.~Forestieri are with
the Telecommunications, Computer Engineering, and Photonics (TeCIP)
Institute, Scuola Superiore Sant'Anna, Pisa, Italy. M.~Secondini,
E.~Forestieri and L.~Pot\`i with the National Laboratory of Photonic
Networks, CNIT, Pisa, Italy. L.~Pot\`i is with the Universitas Mercatorum,
Rome, Italy. Email: stella.civelli@cnr.it.}, \IEEEmembership{Member, IEEE}, Luca Pot\`i, \IEEEmembership{Senior Member, IEEE},
Enrico Forestieri, \IEEEmembership{Life Member, IEEE}, and Marco
Secondini, \IEEEmembership{Senior Member, IEEE}}
\maketitle
\begin{abstract}
Optical satellite uplinks rely on high-power optical amplifiers (HPOAs)
to overcome free-space attenuation and enable long-distance transmission.
However, at high power levels, fiber Kerr nonlinearity becomes significant
and degrades system performance. In this work, we develop a realistic
model for optical uplinks that accounts for nonlinear effects and
analyze their impact, highlighting key differences from conventional
long-haul fiber systems. We then introduce low-complexity digital
signal processing techniques for nonlinearity compensation, based
on constellation shaping via a look-up table (LUT) and a simple nonlinear
phase rotation applied at the transmitter and/or receiver. The LUT
also enables adaptive rate tuning according to channel conditions,
enhancing robustness against link variations. Simulation results show
that the proposed techniques increase the maximum acceptable link
loss by up to 6~dB with negligible complexity. Finally, we show that,
at the system level, propagation in the HPOA can be modeled as a simple
nonlinear phase rotation, equivalent to propagation in a zero-dispersion
noiseless fiber link, and fully characterized by a single parameter---the
characteristic nonlinear power. 
\end{abstract}


\section{Introduction\label{sec:Introduction}}

\IEEEPARstart{R}{ecently}, optical satellite communication has emerged
as an effective technology to complement---and potentially replace---conventional
radio-frequency satellite links \cite{kaushal2016optical,li2022survey,boddeda2023achievableJLT,boddeda2024current}.
Ground-to-satellite optical links, however, suffer from severe impairments
caused by time-varying atmospheric conditions and by the relative
motion between the satellite and the Earth (i.e., varying pointing
angle), both leading to significant attenuation and performance degradation.
To ensure sufficient received power at the satellite, thereby improving
throughput and maintaining resilience under adverse conditions, the
launch power should be as high as possible.

While free-space optical communications provide several advantages
for satellite networks, the ground-to-satellite uplink remains a critical
challenge \cite{Barrios2021}. Its performance is significantly affected
by atmospheric phenomena such as scattering \cite{kaushal2016optical},
molecular absorption \cite{boddeda2023achievableJLT}, and turbulence
\cite{giggenbach2022atmospheric}. Connectivity losses caused by cloud
coverage can be mitigated through the use of optical ground station
diversity \cite{lyras2019cloud}, and adaptive optics techniques are
commonly employed to reduce beam distortions induced by atmospheric
turbulence \cite{roberts2023performance}. However, even with these
mitigation strategies, a fundamental asymmetry persists between the
uplink and downlink power requirements. This asymmetry arises from
the so-called shower curtain effect \cite{dror1998experimental,aviv2006laser},
whereby the impact of a scattering layer depends on its position within
the optical path. Since the ground station is located within the atmospheric
scattering layer, the uplink is more severely degraded than the downlink,
necessitating higher optical transmit power to compensate for atmospheric
impairments. Consequently, scaling the optical power of ground station
transmitters is a key requirement for achieving high data-rate satellite
uplinks.

\begin{figure}
\centering{}\includegraphics[width=1\columnwidth]{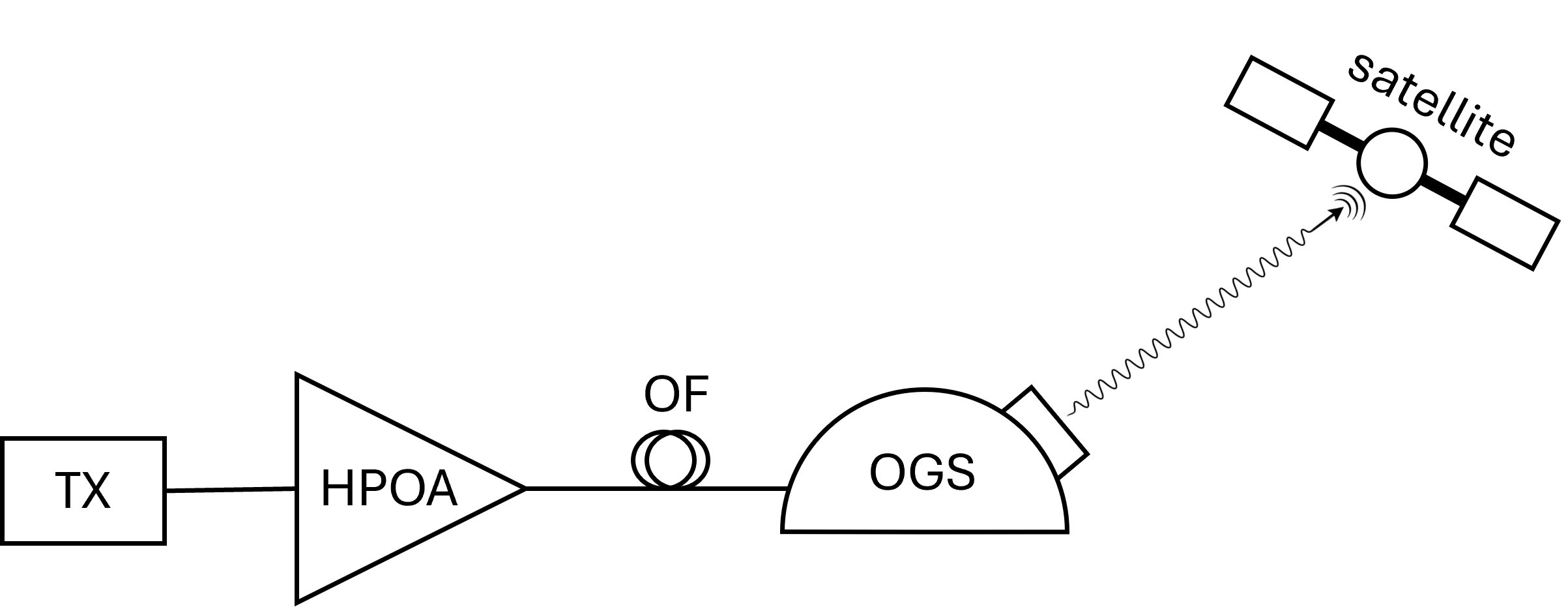}\caption{\label{fig:setup_luca}Ground-to-satellite optical link.}
\end{figure}

A typical architecture for uplink optical transmission is shown in
Fig.\,\ref{fig:setup_luca}, where a standard optical transmitter
(TX) prepares the signal and feeds it into a high-power optical amplifier
(HPOA), which is required to achieve the high optical power needed
for the application. An optical fiber (OF), e.g., an output pigtail
or a short optical patch cord, delivers the signal to the optical
ground station (OGS), which is used for beam preparation and pointing.
In this context, external-cavity lasers followed by fiber-based HPOAs
represent a natural choice for OGS transmitters, owing to their capability
to deliver high average power while maintaining excellent beam quality.
Recently, several HPOAs have been demonstrated, including in-band
core-pumped Er-doped fibers \cite{kotov2022high}, cladding-pumped
Er-doped fibers \cite{kotov2014yb}, and cladding-pumped ErYb co-doped
fibers \cite{matniyaz2020302}. All these approaches exhibit physical
and practical constraints or limitations, making custom designs necessary
depending on the specific application \cite{nicholson2025high}. Indeed,
ongoing research explores more efficient architectures to improve
pump-to-signal conversion and mitigate nonlinear effects, though challenges
in efficiency, thermal stability, and space qualification remain \cite{petrillo2026}.

Among the main challenges, fiber nonlinearities inevitably arise when
high-power signals propagate through optical fibers---either within
the doped fiber inside the amplifier or along the output pigtail and
additional patch cords \cite{Maho2025ampli}. Their mitigation is
essential to increase the launch power, thereby enabling higher throughput
or longer transmission distances \cite{billault2023optical,ciaramella2024nonlinear}.
On one hand, nonlinear effects can be mitigated through dedicated
HPOA designs. For instance, erbium-doped fiber amplifiers (EDFAs)
typically require longer doped fibers compared to erbium-ytterbium
codoped amplifiers (EYDFAs), making the latter more attractive for
reducing nonlinear impairments. However, very-high-power EYDFAs suffer
from higher-order mode power transfer, leading to pointing instability,
as well as long-term power degradation due to photodarkening \cite{nicholson2025high}.
Recently, an Er-based HPOA delivering up to $50$~dBm was proposed
specifically to mitigate nonlinear effects by combining a conventional
EDFA stage with a large-mode-area EDFA stage \cite{grimes202410}.
However, large-mode-area solutions still face practical limitations,
including the limited availability of reliable commercial components
and the difficulty of maintaining stable single-mode operation.

On the other hand, fiber nonlinearity can also be mitigated through
proper signal design and digital signal processing (DSP). For example,
the Kerr nonlinearity arising in a short single-mode-fiber (SMF) pigtail
following an HPOA was investigated in \cite{ciaramella2024nonlinear},
where a compensation method for direct-detection systems was proposed.
Furthermore, coherent optical communications are becoming increasingly
relevant for space applications, enabling more advanced DSP techniques
and higher throughput \cite{li2022survey,guiomar2022coherent}.

In this context, the objective of this work is not to identify optimal
amplifier designs or to provide a comprehensive overview of existing
amplification architectures. Rather, the focus is on modeling nonlinear
effects in coherent optical satellite communication uplinks and on
developing low-complexity DSP-based mitigation strategies that can
be applied independently of the specific amplification solution adopted,
enabling higher launch powers and, consequently, longer transmission
distances or higher data rates. The proposed approaches mainly rely
on the mitigation of nonlinear phase rotation and the containment
of spectral broadening through optimized probabilistic constellation
shaping and pre- and post-compensation techniques. To validate the
proposed model and assess the effectiveness of the considered techniques,
three representative amplifier configurations, characterized by significantly
different performance levels and inspired by both literature \cite{kotov2014yb,grimes202410}
and commercial solutions \cite{keopsys}, are analyzed as case studies.
This work extends the results presented in \cite{civelli2026_satellite_ofc}
by providing a detailed channel model, a theoretical analysis of Kerr
nonlinear effects in HPOAs, and a comprehensive numerical performance
assessment under different compensation techniques and operating scenarios.

The paper is organized as follows. Section II presents a comprehensive
system model and introduces a simplified model that accurately captures
the system behavior. Section III proposes two DSP techniques for the
compensation of nonlinearity: probabilistic constellation shaping
and nonlinear phase compensation (NLPC). Both Sections II and III
provide a comparison with long-haul fiber systems, whose nonlinear
effects have been thoroughly investigated over the past decades. Section
IV presents the system setup and a detailed numerical analysis of
the system and the proposed DSP techniques under different conditions.
Finally, Section V draws the conclusions. 

\section{System Model\label{sec:System-Model}}

\begin{figure*}
\centering{}\includegraphics[width=2\columnwidth]{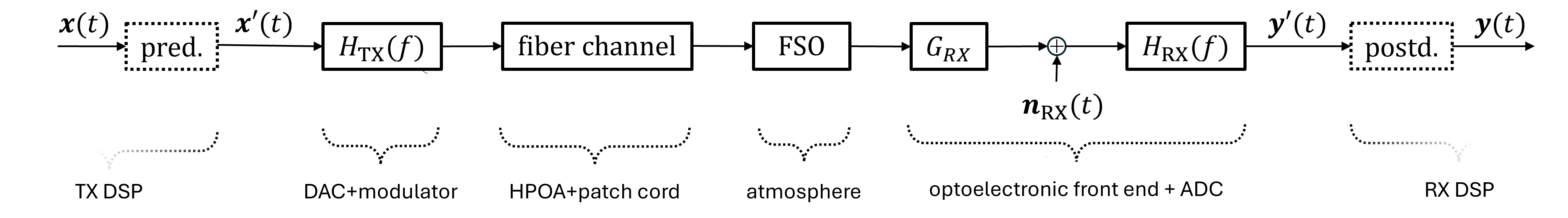}\caption{\label{fig:sysmodel}Equivalent baseband model of the coherent optical
uplink system, including hardware and channel impairments. Dashed
blocks denote optional DSP compensation functions at the TX and RX.}
\end{figure*}
We consider a coherent ground-to-satellite optical link, sketched
in Fig.~\ref{fig:setup_luca}. Fig.~\ref{fig:sysmodel} illustrates
the equivalent complex baseband representation adopted throughout
this work. The system comprises a ground-based transmitter, a short
high-power optical fiber stage including amplification, a free-space
optical (FSO) link toward the satellite, and a coherent receiver.
In the following, each subsystem is described in detail.

\subsection{Transmitter}

At the transmitter (TX), information bits are mapped by DSP onto
a dual-polarization (DP) $M$-ary quadrature amplitude modulation
(QAM) signal with baud rate $R$. The resulting complex baseband signal
can be written as
\begin{equation}
\mathbf{x}(t)=\sum_{k}\mathbf{a}_{k}\,p(t-kT)\label{eq:TX_signal}
\end{equation}
where $\mathbf{a}_{k}=[a_{k,x},a_{k,y}]^{\mathrm{T}}$ contains the
couple of complex QAM symbols transmitted on the two orthogonal polarizations,
$p(t)$ is the pulse shape, and $T=1/R$ the symbol time. Optional
digital pre-distortion may be applied at this stage.

Each polarization component is decomposed into its in-phase and quadrature
parts. Consequently, the DP signal is represented by four real-valued
electrical waveforms (I/Q components for the two polarizations), which
are converted into analog form by four digital-to-analog converters
(DACs).

The resulting analog signals are applied to a dual-polarization IQ
modulator, which maps them onto the two orthogonal polarization states
of a continuous-wave optical carrier generated by a narrow-linewidth
laser source. Under the complex baseband representation adopted in
this work, the combined effect of DACs, drivers, and modulator bandwidth
limitations is captured by $H_{\mathrm{TX}}(f)$.

The resulting optical signal is subsequently amplified by an HPOA
and delivered to the OGS output through a short fiber section (patch
cord or pigtail).

\subsection{Optical Fiber Channel}

The optical signal propagates through two short fiber sections: the
active fiber inside the HPOA and the passive delivery fiber (patch
cord). Although their total length is typically below 100~m, the
launched optical power may exceed 35--40\,dBm, making nonlinear
effects significant.

Propagation of the dual-polarization signal in both sections is described
by the Manakov equation for the complex envelope $\mathbf{u}(z,t)=[u_{x}(z,t),u_{y}(z,t)]^{T}$,
normalized such that $E\{\Vert\mathbf{u}(z,t)\Vert^{2}\}=1$ \cite{menyuk2006interaction}\footnote{The small random birefringence unavoidably present in optical fibers
gives rise to an additional polarization-dependent nonlinear term.
In standard telecom fibers, the birefringence correlation length is
typically much shorter than the propagation distance. As a result,
the contribution of this term averages out along the link and is generally
negligible at the system level. Its possible relevance in the scenario
considered here is left for future investigation. Other nonlinear
mechanism (e.g., Raman, Brillouin, ...) are not included in the model,
since Kerr effect is expected to be the dominant impairment \cite{ciaramella2024nonlinear}.}
\begin{equation}
\frac{\partial\mathbf{u}}{\partial z}=j\frac{\beta_{2}}{2}\frac{\partial^{2}\mathbf{u}}{\partial t^{2}}-j\gamma Pg(z)\Vert\mathbf{u}\Vert^{2}\mathbf{u}+\mathbf{n}_{\mathrm{ASE}}\label{eq:Manakov}
\end{equation}
where $P$ is the output optical power (on both polarizations), $\gamma$
is the Kerr nonlinear coefficient, $\beta_{2}$ is the group-velocity
dispersion, $g(z)$ is the normalized longitudinal power profile ($g(L)=1$,
with $L$ the total fiber length) accounting for both gain and loss,
and $\mathbf{n}_{\mathrm{ASE}}(z,t)$ is the amplified spontaneous
emission (ASE) noise. Below, the dominant physical effects are discussed.

\subsubsection{Amplification and Noise}

The HPOA is modeled as a distributed-gain active fiber generating
ASE along the propagation coordinate $z$. The longitudinal power
profile $g(z)$ and the power spectral density (PSD) of the ASE noise
term $\mathbf{n}_{\mathrm{ASE}}(z,t)$ are obtained from the standard
coupled propagation and population equations governing doped-fiber
amplifiers, evaluated under the assumption of a continuous-wave input
signal, flat gain and ASE spectra over the signal bandwidth. Under
these assumptions, the computed gain and noise profiles are treated
as known quantities and incorporated into the normalized Manakov equation
(\ref{eq:Manakov}) to assess their impact on signal propagation.
By contrast, the patch cord is a passive fiber, where $g(z)$ is determined
only by fiber attenuation and no ASE noise is generated ($\mathbf{n}_{\mathrm{ASE}}(z,t)=0$).

As will be shown later, although gain and ASE are generated continuously
along the HPOA, their detailed longitudinal distribution is irrelevant
from a system-level perspective, since only the output quantities
determine performance. To characterize them, we introduce two lumped
parameters. The first is the amplifier gain $G_{\mathrm{HPOA}}=P/P_{\mathrm{IN}}$,
defined as the ratio between output and input signal power. The second
is the noise figure $F_{\mathrm{HPOA}}$, which quantifies the degradation
of the signal-to-noise ratio (SNR) introduced by the amplifier and
therefore determines the PSD of the ASE noise accumulated at the HPOA
output, $N_{\mathrm{HPOA}}=G_{\mathrm{HPOA}}F_{\mathrm{HPOA}}h\nu/2$
(per polarization).

\subsubsection{Chromatic Dispersion}

The first term of the right-hand side of (\ref{eq:Manakov}) is responsible
for chromatic dispersion, which induces a temporal broadening of the
propagating optical pulses. For a pulse of width $T_{0}$, the amount
of broadening increases with the ratio $L/L_{D}$, where

\begin{equation}
L_{D}\triangleq\frac{T_{0}^{2}}{|\beta_{2}|}
\end{equation}
is the dispersion length \cite[eq (3.1.5)]{agrawalNL}. We can readily
verify that in the typical scenario considered in this work, $L\ll L_{D}$,
indicating a negligible broadening of the propagating signal. For
instance, for a 100~GBd signal in a standard SMF, assuming $\beta_{2}=-\unit[21.7]{ps^{2}/km}$
and $T_{0}\approx T=\unit[10]{ps}$, $L_{D}\approx\unit[4.6]{km}$,
whereas the total fiber length $L$ is typically below 100~m. 

\subsubsection{Kerr Nonlinearity}

In contrast, Kerr nonlinearity---the second term on the right-hand
side of (\ref{eq:Manakov})---plays a central role due to the high
optical power. When $L\ll L_{D}$, chromatic dispersion can be neglected
and, in the absence of ASE noise, propagation is dominated by the
Kerr nonlinear term. In this case, (\ref{eq:Manakov}) can be solved
in closed form as
\begin{equation}
\mathbf{u}(L,t)=\mathbf{u}(0,t)\exp\left(-j\bar{\phi}\Vert\mathbf{u}(0,t)\Vert^{2}\right)\label{eq:NLzerodisp}
\end{equation}
with
\begin{equation}
\bar{\phi}=P\int_{0}^{L}\gamma g(z)dz\label{eq:average_NLPR}
\end{equation}
According to (\ref{eq:NLzerodisp}), the optical signal undergoes
a phase rotation proportional to its instantaneous power, a phenomenon
known as \emph{self-phase modulation} (SPM). The strength of SPM depends
on the dimensionless parameter $\bar{\phi}$ in (\ref{eq:average_NLPR}),
which represents the average accumulated nonlinear phase rotation
(NLPR). For convenience, we rewrite it as $\bar{\phi}=P/P_{\mathrm{NL}}$,
where we have introduced the characteristic nonlinear power
\begin{equation}
P_{\mathrm{NL}}=\left(\int_{0}^{L}\gamma g(z)dz\right)^{-1}\label{eq:characteristic_NL_power}
\end{equation}
representing the power scale at which SPM becomes significant. For
instance, the EDFA considered in Section~\ref{subsec:System-setup}
(Setup~A) yields $P_{\text{NL}}\approx\unit[42.7]{dBm}$.

We distinguish three main effects associated with SPM.

\paragraph{Nonlinear phase rotation (NLPR)}

Each received symbol undergoes a different power-dependent phase rotation.
In the absence of chromatic dispersion, symbols do not spread nor
interact through linear memory, and therefore the nonlinear distortion
remains temporally uncorrelated. In principle, NLPR is a deterministic
effect which can be exactly compensated by applying the inverse phase
rotation at either the transmitter or the receiver, as discussed in
Section~\ref{subsec:Nonlinear-phase-compensation}. In practice,
however, compensation is fundamentally limited by the additional effects
described below. Moreover, unlike long-haul dispersive links, the
absence of temporal correlation in the NLPR prevents its mitigation
through conventional carrier phase recovery algorithms that rely on
temporal averaging \cite{civelli2022JLTBPS}.

\paragraph{Spectral broadening}

SPM inherently implies spectral broadening, since the phase shift
in (\ref{eq:NLzerodisp}) induces a time-varying frequency chirp that
depends on the derivative of the signal intensity. In conventional
systems operating on standard SMF fibers in the C band, spectral broadening
is strongly suppressed by chromatic dispersion  and is therefore
typically negligible. By contrast, in the present scenario, where
$L\ll L_{D}$, it may become significant and, combined with the finite
bandwidth of practical TXs and RXs, degrade system performance and
limit the effectiveness of NLPR compensation.

In this dispersionless regime, under the assumption of independent
and statistically equivalent polarizations with circularly symmetric
complex Gaussian statistics, the evolution of the PSD can be derived
analytically. Letting 
\begin{equation}
R(z,\tau)\triangleq E\{u_{x}(z,t+\tau)u_{x}^{*}(z,t)\}=E\{u_{y}(z,t+\tau)u_{y}^{*}(z,t)\}
\end{equation}
denote the autocorrelation function of each polarization component
of the normalized envelope at distance $z$, one obtains (see Appendix)
\begin{equation}
R(L,\tau)=\frac{R(0,\tau)}{\left[1+\bar{\phi}_{\mathrm{NL}}^{2}(1/4-|R(0,\tau)|^{2})\right]^{3}}\label{eq:autocrrelation_evolution}
\end{equation}
The PSD follows from the Wiener--Khinchin theorem as 
\begin{equation}
S(z,f)=\int_{-\infty}^{\infty}R(z,\tau)e^{-j2\pi f\tau}d\tau.
\end{equation}
The accuracy of this analytical expression will be verified in Section~\ref{subsec:Spectral-broadening_results}.

\paragraph{Signal--noise interaction}

During propagation inside the HPOA, ASE noise is continuously generated,
as described by the distributed noise term in (\ref{eq:Manakov}).
In principle, this ASE component co-propagates with the signal and
causes random intensity fluctuations that, through (\ref{eq:NLzerodisp}),
are converted to random phase fluctuations (nonlinear phase noise).

In the considered scenario, however, the amount of ASE generated within
the amplifier is typically negligible compared to other sources of
noise in the system. As a consequence, the corresponding nonlinear
signal--noise interaction is also negligible, and the overall SNR
is essentially determined by the receiver noise.

On the other hand, receiver noise, although it does not physically
interact with the signal inside the nonlinear fiber, may give rise
to an indirect but analogous form of signal--noise interaction when
attempting to invert (\ref{eq:NLzerodisp}) by DSP at the receiver.
This mechanism, which will be analyzed in the following, ultimately
limits the practical mitigation of NLPR.

\subsection{Free-Space Optical Channel}

After leaving the OGS, the optical signal propagates through an FSO
link toward the satellite.

All time-varying propagation effects---such as atmospheric turbulence,
scattering, and molecular absorption---are treated in a static manner.
The wireless channel is therefore simply modeled as a constant attenuation
factor $\mathcal{L}$, representing the overall link loss between
ground and satellite.

\subsection{Receiver}

At the receiver (RX), a preamplified coherent opto-electronic front-end
is considered, comprising an optical pre-amplifier with gain $G_{\mathrm{RX}}$
and noise figure $F_{\mathrm{RX}}$; a polarization- and phase-diversity
coherent receiver with balanced photodetectors; and four analog-to-digital
converters (ADCs), one per each quadrature component of each polarization.
The optical pre-amplifier introduces an additive white Gaussian noise
(AWGN) contribution $\mathbf{n}_{\mathrm{RX}}(t)$ with PSD $N_{\mathrm{RX}}=G_{\mathrm{RX}}F_{\mathrm{RX}}h\nu/2$
per polarization. The receiver is assumed ideal, except for the finite
electrical bandwidth of the photodetectors and ADCs, by the low-pass
transfer function $H_{\mathrm{RX}}(f)$.

The digitized samples are finally processed by DSP, which includes
optional NLPR compensation, matched filtering, and symbol-rate sampling.
More details about DSP are provided in Section~\ref{sec:Digital-Techniques-for}
and \ref{subsec:System-setup}.

\subsection{Simplified Model\label{subsec:Simplified-Model}}

\begin{figure}
\centering{}\includegraphics[width=1\columnwidth]{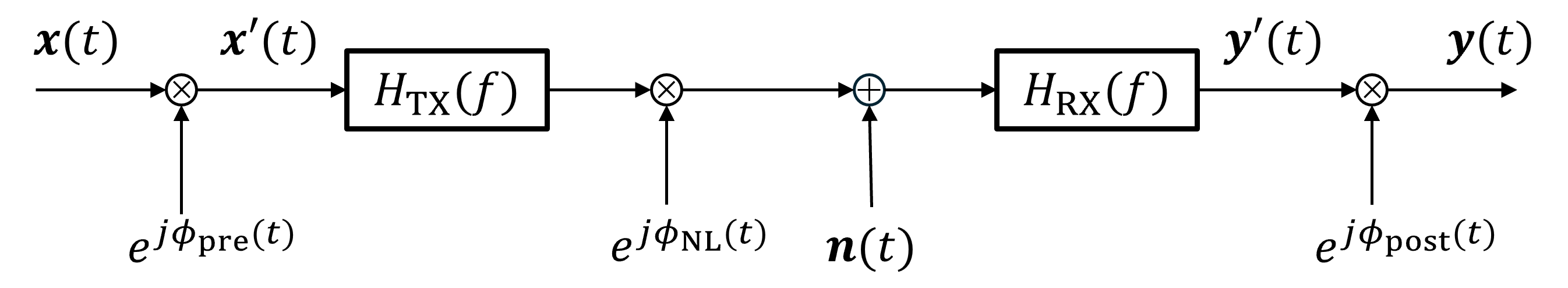}\caption{\label{fig:figSimpleModel}Simplified equivalent baseband model derived
from Fig.~\ref{fig:sysmodel} under the approximations in Section~\ref{subsec:Simplified-Model},
used for theoretical analysis and as a guideline for the development
of the proposed DSP compensation techniques.}
\end{figure}

Based on the above considerations, the baseband model in Fig.~\ref{fig:sysmodel}
can be simplified as in Fig.~\ref{fig:figSimpleModel}, where the
fiber channel is simply modelled by the NLPR $\phi_{\mathrm{NL}}(t)=-\bar{\phi}\Vert\mathbf{u}(0,t)\Vert^{2}$
in (\ref{eq:NLzerodisp}) (neglecting dispersion and signal--noise
interaction), and the AWGN term $\mathbf{n}(t)=[n_{x}(t),n_{y}(t)]^{T}$
accounts for the contribution of all noise sources and for the effect
of gains and losses, with an overall SNR (per symbol) 
\begin{equation}
\mathrm{SNR}=\frac{P}{R_{s}h\nu(\mathcal{L}F_{\mathrm{RX}}+G_{\mathrm{HPOA}}F_{\mathrm{HPOA}})}\approx\frac{P}{R_{s}h\nu\mathcal{L}F_{\mathrm{RX}}}\label{eq:SNR}
\end{equation}
The approximation in (\ref{eq:SNR}) holds when $\mathcal{L}\gg G_{\mathrm{HPOA}}$,
i.e., in the practically relevant scenario of strong channel loss.

This simplified model captures the most relevant characteristics of
the channel, namely, AWGN, bandwidth constraints, NLPR, and the associated
spectral broadening. It will therefore be used in the following to
design and optimize the system and to provide a theoretical interpretation
of the results.

In particular, based on this model, we restrict transmitter predistortion
and receiver post-compensation to simple phase rotations---$\phi_{\mathrm{pre}}(t)$
and $\phi_{\mathrm{post}}(t)$, respectively---as detailed in Section~\ref{subsec:Nonlinear-phase-compensation}.

On the other hand, simulation results will be obtained using the more
accurate physical model described previously. Nevertheless, we will
demonstrate that the simplified model reproduces them with good accuracy
and can, in principle, replace the full physical model even for reliable
performance estimation.

\section{Digital Techniques for Nonlinearity Compensation\label{sec:Digital-Techniques-for}}

In this section, we discuss and analyze DSP techniques for the mitigation
of the HPOA nonlinearity. The proposed techniques are inherited from
optical fiber communication and tailored to the HPOA scenario.

\subsection{Probabilistic Constellation Shaping}

\begin{figure}
\begin{centering}
\includegraphics[width=0.5\columnwidth]{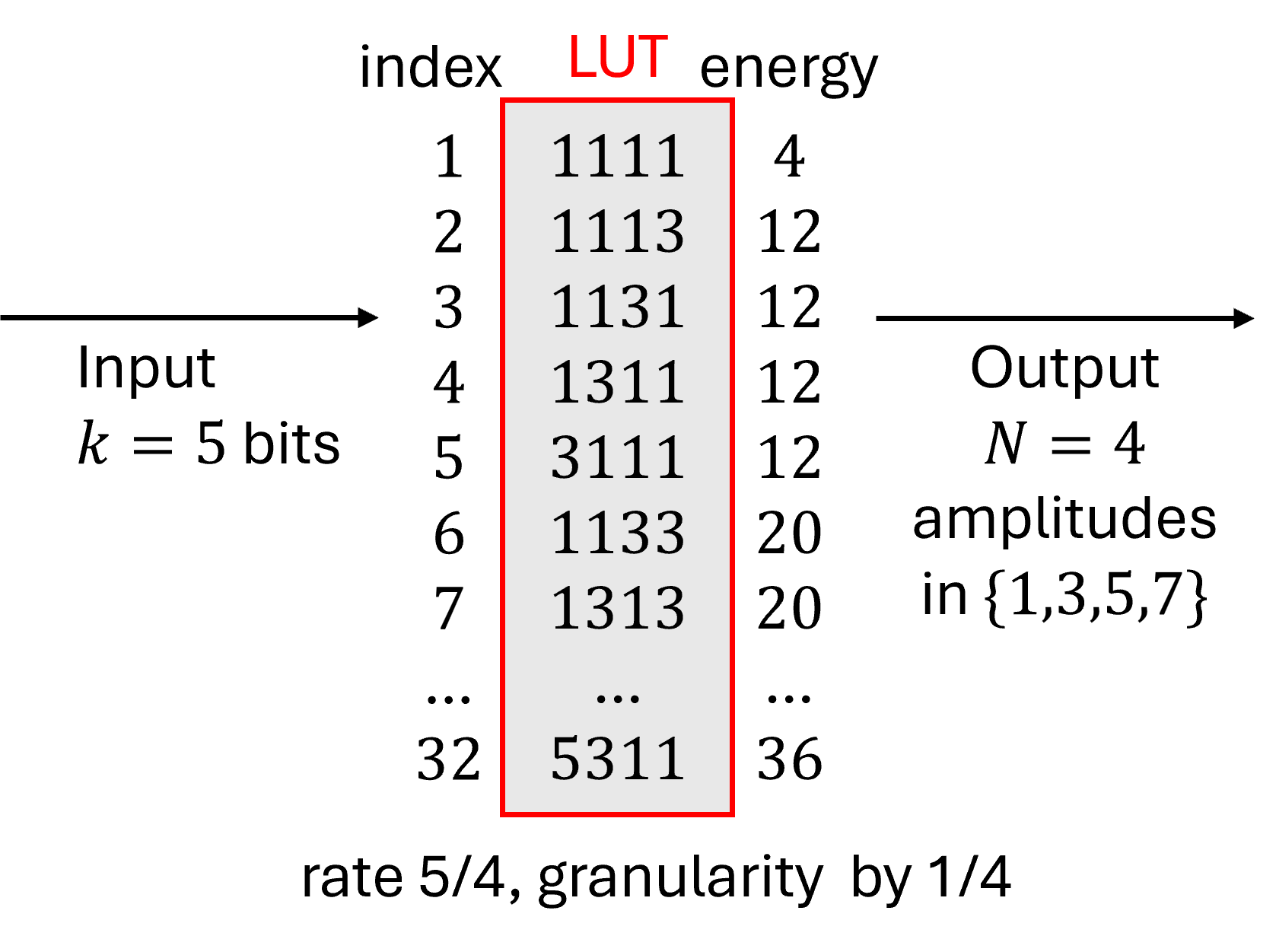}
\par\end{centering}
\caption{\label{fig:lut45}LUT with $N=4$ amplitudes from $\{1,3,5,7\}$ and
$32$ entries $(k=5)$, corresponding to a $64$QAM SpSh constellation
with rate $4.5$\,bits/2D.}
\end{figure}

Probabilistic constellation shaping has been known for several decades,
and nowadays is commonly adopted in optical fiber coherent systems
with the dual objective of improving system performance and enabling
fine-tuning of the transmission rate. Probabilistic shaping is commonly
implemented through the probabilistic amplitude shaping (PAS) architecture,
in which a distribution matcher (DM) maps information bits onto amplitudes
with the desired probability distribution, while the signs remain
uniformly distributed and are used for forward error correction (FEC)
\cite{bocherer2015bandwidth}. The main component of PAS is the DM,
which maps $k$ information bits onto $N$ shaped amplitudes (where
$N$ denotes the block length) drawn from the $M$-ary alphabet $\{1,3,\dots,2M-1\}$,
achieving a DM rate $R_{\text{DM}}=k/N$. The $N$ amplitudes are
then combined with $N$ signs (with uniform distribution), to form
$N/4$ couples of complex QAM symbols modulating the signal in (\ref{eq:TX_signal}).\footnote{Different mapping strategies are possible---e.g., mapping the DM
amplitudes independently on each quadrature component of each polarization
\cite{civelli2022JLTBPS}---but are not considered in this work.} The accuracy of the DM depends on the specific implementation but
typically improves when $N$ increases \cite{schulte2016CCDM,gultekin2018Sphereshaping,civelli2022JLTBPS}.
In general, algorithmic DMs, such as constant-composition DM (CCDM),
enumerative sphere shaping (ESS), and their variants, achieve good
performance at the cost of non-negligible complexity for large $N$
\cite{schulte2016CCDM,gultekin2018Sphereshaping}. By contrast hierarchical
DMs based on look-up-tables (LUTs) have negligible complexity but
reduced performance \cite{yoshida2019hierarchicalDM,civelli2020entropy}.
When $N$ is very small, the DM can be implemented efficiently with
a single LUT, simply storing the $2^{k}$ best sequences of $N$ amplitudes.
In the linear regime with an average power constraint, the optimal
DM scheme is denoted as sphere shaping (SpSh), corresponding to mapping
information on sequences with minimum energy. An example of LUT-based
SpSh encoding is shown in Fig.\,\ref{fig:lut45} for $k=5$, $N=4$,
and $M=4$, ($R_{\mathrm{DM}}=1.25$\,bits/amplitude, 64QAM constellation).
The $k$ input bits determine the LUT address containing the output
amplitude sequence.  Correspondingly, the decoding LUT has $M^{N}$
entries---one for each possible sequence of $N$ amplitudes---each
returning the associated $k$ bits.\footnote{Since not all $M^{N}$ amplitude sequences are actually needed, a
more efficient implementation of the decoding LUT may be devised.} Overall, the memory required for the encoding and decoding LUTs is
$2^{k}N\log_{2}M$ and $M^{N}k$ bits, respectively.

PAS provides two main advantages, discussed below: shaping gain and
rate adaptability. Thanks to these features, PAS has also been proposed
for free-space optical communications \cite{guiomar2022coherent}.

Rate adaptability refers to the ability to vary the information rate
(IR) for a fixed FEC code rate $c$ and constellation order, by solely
adjusting the DM rate. Specifically, the information rate is given
by $\text{IR}=R_{\text{DM}}+1-(1-c)m$ in bits per 1D symbol, where
$m=\log_{2}M+1$ is the number of bits per 1D symbol \cite{buchali2016JLT}.
This property is particularly attractive for satellite communications,
where channel conditions can vary significantly, for instance due
to cloud coverage. When using a LUT with block length $N$, the rate
can be adapted by simply decreasing $k$---i.e., using only the first
$2^{k}$ entries of the LUT---in steps of $1/N$, with finer granularity
for larger $N$.

Shaping gain refers to the improved performance compared to uniform
constellations. In the linear regime, PAS aims at reducing the mean
energy per symbol for a fixed IR, and approaches Shannon capacity
as $N$ increases, providing up to 1.53\,dB gain over an AWGN channel
compared to the uniform constellation by targeting the Maxwell--Boltzmann
(MB) distribution, a sort of \emph{discretized} Gaussian distribution
\cite{bocherer2015bandwidth}.

PAS has also been investigated for mitigating nonlinearities in optical
fiber communication, but with limited gains \cite{fehenberger2016JLT,gultekin2018Sphereshaping,civelli2022JLTBPS}.
Here the mechanism relies on the correlations induced by the DM among
the amplitudes within each block, which reduce the nonlinear interference
generated during propagation. This effect is more pronounced for shorter
block lengths $N$. As a result, the overall shaping gain is maximized
for a block length that balances linear and nonlinear effects and
depends on the channel memory, i.e., on the accumulated dispersion.
The optimal $N$ is generally very long ($N$>100) for multi-span
fiber links (in C-band with SMF), and shorter ($N\sim30$) for single-span
links, but in any case too long for a single-LUT implementation. Moreover,
the long memory of these fiber channels implies that carrier phase
recovery algorithms that are commonly employed to address laser phase
noise are also capable of mitigating the generated nonlinear interference,
hiding the nonlinear shaping gain in most cases \cite{civelli2022JLTBPS,civelli2024sequenceJLT}.

The picture is quite different for satellite optical coherent up-links
with HPOAs. While the behavior in the linear regime remains unchanged,
the nonlinear regime differs due to the nearly dispersionless nature
of the fiber channel, which makes nonlinearity an almost instantaneous
(memoryless) effect, as discussed in Section~\ref{sec:System-Model}.
This leads to two key consequences, which will be quantified and analyzed
in Section~\ref{subsec:System-performance-and}: i) the optimal PAS
block length $N$ becomes much shorter, making a single-LUT implementation
of the DM---such as the one depicted in Fig. \ref{fig:lut45}---particularly
attractive; ii) carrier phase recovery algorithms are ineffective
in mitigating nonlinear interference, increasing the relevance of
the nonlinear shaping gain provided by PAS.

\subsection{Nonlinear Phase Compensation (NLPC)\label{subsec:Nonlinear-phase-compensation}}

Assuming the simplified model in Fig.~\ref{fig:figSimpleModel},
fiber propagation reduces to the phase rotation in (\ref{eq:NLzerodisp}),
which can be inverted at the TX or RX DSP by applying the inverse
rotation. We refer to this operation as \emph{nonlinear phase compensation}
(NLPC). In the absence of bandwidth limitations and noise, exact inversion
could be equivalently performed at the TX or RX. In practice, however,
bandwidth limitations and noise prevent exact inversion, as discussed
below.

At the RX, the transfer function $H_{\mathrm{RX}}(f)$ filters out
part of the spectrally broadened signal, preventing exact inversion
of (\ref{eq:NLzerodisp}). Similarly, at the TX, accurate pre-compensation
would require generating frequency components that may lie outside
the available bandwidth defined by $H_{\mathrm{TX}}(f)$. To mitigate
the effect of bandwidth limitations, it would be advantageous to split
the NLPC equally between TX and RX, minimizing the maximum spectral
broadening experienced by the signal at both locations. However, the
noise injected between the HPOA and the RX breaks the symmetry between
TX and RX processing, degrading only the effectiveness of RX-side
NLPC. Furthermore, complexity and power consumption constraints are
also asymmetric between satellite and ground terminals, generally
favoring TX processing.

For these reasons, we consider a split NLPC between TX and RX, with
splitting ratio $\kappa$
\begin{align}
\mathbf{x}'[k] & =\mathbf{x}[k]\exp(j\kappa\bar{\phi}\Vert\mathbf{x}[k]\Vert^{2})\label{eq:NLPC-TX}\\
\mathbf{y}[k] & =\mathbf{y}'[k]\exp(j(1-\kappa)\bar{\phi}\Vert\mathbf{y}'[k]\Vert^{2})\label{eq:NLPC-RX}
\end{align}
where $\mathbf{x}[k]$, $\mathbf{x}'[k]$, $\mathbf{y}'[k]$, and
$\mathbf{y}[k]$ denote the transmitted signal (before and after NLPC)
and the received signal (before and after NLPC), respectively, as
shown in Fig.~\ref{fig:sysmodel} and \ref{fig:figSimpleModel},
sampled at $n$ samples per symbol. Based on the above discussion,
the splitting ratio $\kappa$ can be either optimized to maximize
performance or set to $\kappa=1$, corresponding to NLPC entirely
performed at the TX. Both cases are analyzed in Section~\ref{subsec:System-performance-and}.

The complexity of NLPC, at the TX or RX side, measured as number of
real multiplications (RM) per 2D symbols, is 
\begin{equation}
C_{\text{NLC}}=5.5n\,\,\text{RM}/2\text{D}\label{eq:complexityNLC}
\end{equation}
assuming that each complex multiplication is implemented with $3$
real multiplications \cite{wenzler1995new}.

\section{Numerical Results}

\subsection{System Setup\label{subsec:System-setup}}

\begin{figure}
\centering{}\includegraphics[width=1\columnwidth]{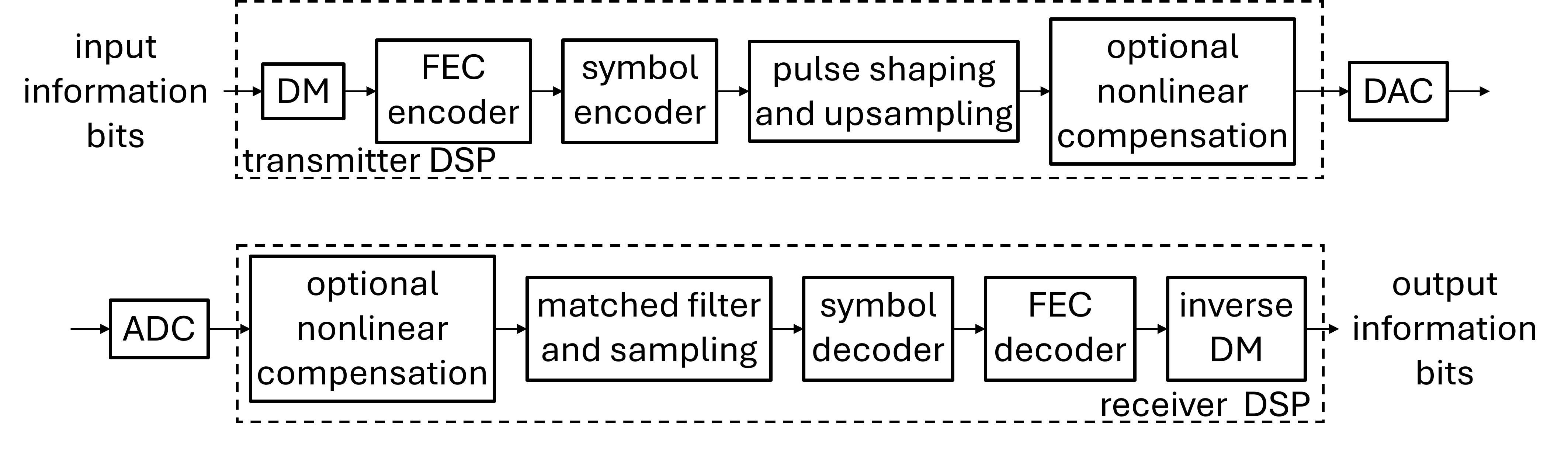}\caption{\label{fig:figDSP}Transmitter and receiver DSP.}
\end{figure}

System performance is evaluated through numerical simulations. A detailed
system model is provided in Section\,\ref{sec:System-Model}, while
the transmitter and receiver DSP chains---operating with $n$ samples/symbol---are
shown in Fig.\,\ref{fig:figDSP} and described below.

The TX-DSP, located at the ground station, maps information bits on
symbols $\mathbf{a}_{k}$ in (\ref{eq:TX_signal}) drawn from a DP
QAM constellation with uniform or probabilistically shaped distribution.
In the latter case, the system uses the PAS approach, which combines
DM and FEC in reverse concatenation order (the latter not included
in our simulations) \cite{bocherer2015bandwidth}. Unless otherwise
stated, the pulse shape $p(t)$ in (\ref{eq:TX_signal}) has a root-raised
cosine spectrum with roll-off $0.05$, and the symbol rate is $R=100$\,GBd.
An optional NLPC stage, defined in (\ref{eq:NLPC-TX}) and implemented
with oversampling factor $n$, may be applied to (\ref{eq:TX_signal})
to obtain the signal $\mathbf{x}'(t)$.

Three different optical amplification configurations are considered:
\begin{lyxlist}{00.00.0000}
\item [{Setup~A:}] an EDFA with $33$\,m of Er-doped fiber (dispersion
$\beta_{2}=\unit[-21.7]{ps^{2}/km}$, nonlinear coefficient $\gamma=\unit[3.6]{W^{-1}km^{-1}}$),
resulting in an overall characteristic nonlinear power $P_{\mathrm{NL}}\approx\unit[42.68]{dBm}$;
\item [{Setup~B:}] the 100\,W HPOA reported in \cite{kotov2014yb}, consisting
of $40$\,m of large-mode-area Er-doped fiber ($\beta_{2}=\unit[-21.7]{ps^{2}/km}$,
$\gamma=\unit[0.2]{W^{-1}km^{-1}}$), followed by $3$\,m of standard
SMF ($\beta_{2}=\unit[-21.7]{ps^{2}/km}$, $\gamma=\unit[1.27]{W^{-1}km^{-1}}$)
used as pigtails and patch cords, yielding $P_{\text{NL}}=\unit[51.24]{dBm}$;
\item [{Setup~C:}] the master-oscillator power-amplifier scheme described
in \cite{matniyaz2020302}, capable of delivering a record output
power of 302\,W, where the power amplifier consists of 6\,m of Er/Yb
co-doped large-mode-area fiber ($\beta_{2}=\unit[-21.7]{ps^{2}/km}$,
$\gamma=\unit[1.27]{W^{-1}km^{-1}}$). To emulate a non-co-located
optical transmitter and telescope architecture, an additional 20\,m
of passive fiber, with propagation characteristics identical to those
of the active fiber, is included at the amplifier output to represent
pigtails and interconnecting patch cords, resulting in an overall
$P_{\mathrm{NL}}=\unit[53.37]{dBm}$.
\end{lyxlist}
Unless otherwise specified, the results presented in the following
are based on Setup A. A comparison with Setups B and C, which exhibit
improved performance, as well as a validation of the simplified model
across all three configurations, is provided in the final part of
the paper. Fiber propagation according to (\ref{eq:Manakov}) is evaluated
numerically using the split-step Fourier method (SSFM) with a sufficiently
large number of steps, including dispersion, nonlinear Kerr effects,
gain, and distributed noise due to ASE.

After the HPOA, the signal is launched into free space and received
at the satellite station. Propagation through the FSO channel is simply
modeled as a lumped attenuation $\mathcal{L}$. The noise figure of
the RX-side EDFA is $F_{\text{RX}}=\unit[4]{dB}$. The limited bandwidth
of both TX- and RX-side opto-electronics components (DAC, ADC, modulator
and photoreceiver) is accounted for by ideal rectangular filters $H_{\mathrm{TX}}(f)=H_{\mathrm{RX}}(f)$
with lowpass bandwidth $B=55$\,GHz. The overall signal bandwidth,
including the pulse-shaping roll-off, is $1.05\cdot R=\unit[52.5]{GHz}$,
leaving a small margin (about 5\%) for spectral broadening. At the
RX-DSP, an optional NLPC---defined in (\ref{eq:NLPC-RX}) and implemented
with oversampling factor $n$ as at the TX-side---can be applied,
followed by matched filter and sampling at symbol time $1/R$. The
resulting samples---a noisy version of the transmitted symbols---are
compared with the transmitted symbols to evaluate the generalized
mutual information (GMI), which quantifies the achievable information
rate with ideal FEC and bit-wise decoding \cite{alvarado2018achievable,fehenberger2018multiset}.
In our simulations, the inverse DM and FEC decoding shown in Fig.\,\ref{fig:figDSP}
are not applied, since they are not required to compute the GMI. The
TX and LO lasers are assumed ideal (zero linewidth), as the phase
noise of typical telecom lasers at these baud rates can be effectively
compensated by standard carrier phase recovery algorithms without
performance penalty \cite{pfau2009BPS}. For a given launch power
$P$, the performance is measured in terms of \emph{acceptable link
loss}, which is the maximum loss $\mathcal{L}$ for which a desired
GMI can be achieved. Our metric is analogous to the power budget---typically
defined as the maximum allowable loss for a desired pre-FEC bit error
rate (BER)---but replaces the pre-FEC BER threshold with a target
GMI. Achieving the target GMI (measured in bits/2D) ensures that,
with an ideal FEC, the system can operate with arbitrarily low error
probability at the corresponding bit rate $R_{b}=2R\cdot\mathrm{GMI}$,
where the factor 2 accounts for the two polarizations.\footnote{For a specific realistic FEC, the target GMI can be replaced by the
normalized GMI threshold evaluated for that FEC scheme \cite{cho2017normalized}.} In our setup, unless otherwise stated, we consider two cases: (i)
a target GMI of 3\,bits/2D (corresponding to an achievable bit rate
$R_{b}=\unit[600]{Gb/s}$) on the shaped 64QAM constellation with
shaping rate 4.5\,bits/2D or with uniform (U)-16QAM or 64QAM; and
(ii) target GMI of 5\,bits/2D (corresponding to $R_{b}=\unit[1]{Tb/s}$)
on the 256QAM constellation with shaping rate 6.5\,bits/2D or U-64QAM;
the shaping rate has been optimized offline. Clearly, the acceptable
link loss depends on the launched power $P$. It increases linearly
with $P$ until fiber nonlinearity remains negligible ($P\ll P_{\mathrm{NL}}$),
then reaches a peak at the optimal power and decreases afterwards
due to the NLPR and spectral broadening induced by the nonlinearity.
Accordingly, we define the \emph{maximum acceptable link loss} as
the peak value achieved at this optimal launch power.

\subsection{Spectral Broadening\label{subsec:Spectral-broadening_results}}

\begin{figure}
\begin{centering}
\includegraphics[width=1\columnwidth]{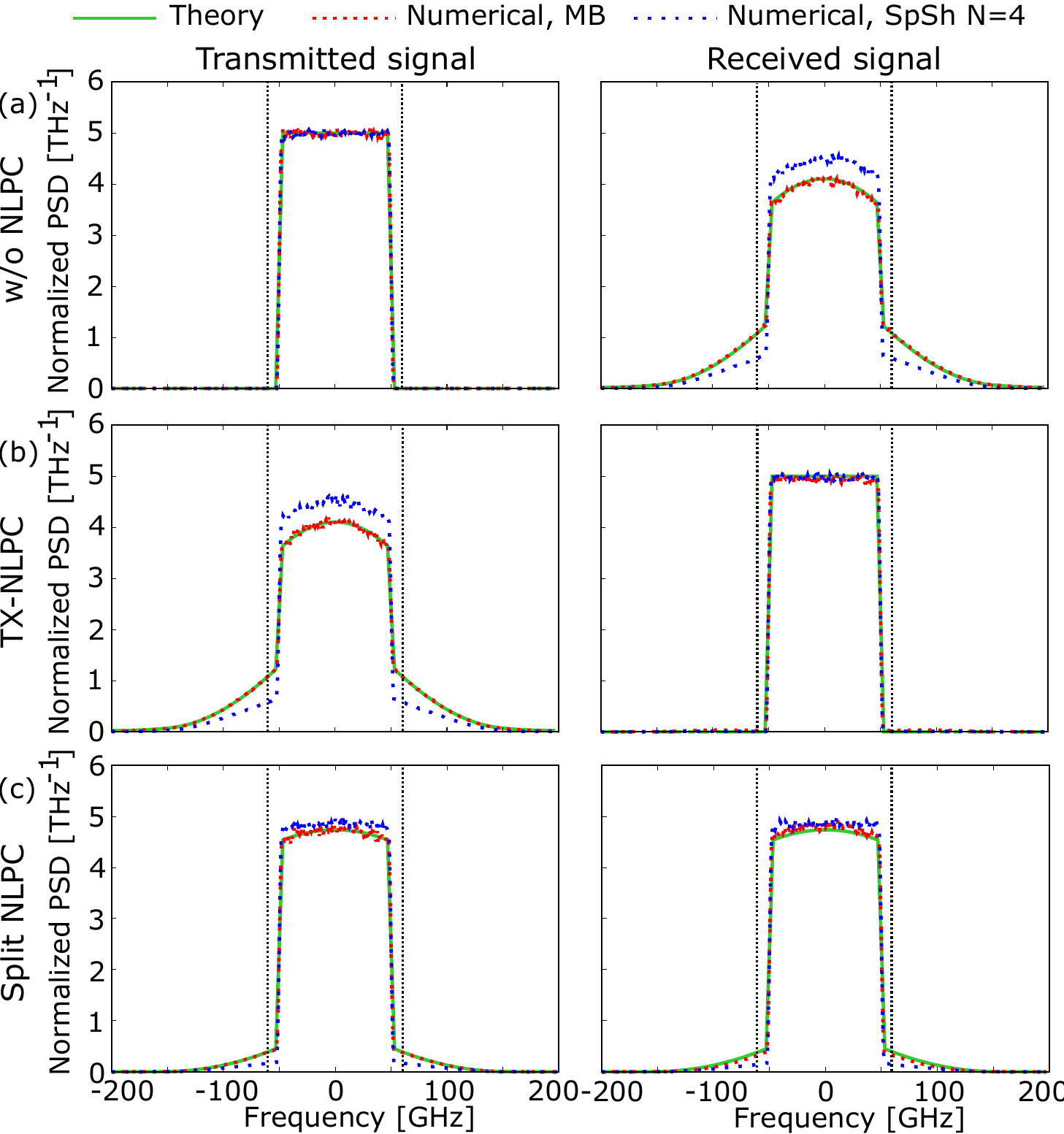}
\par\end{centering}
\caption{\label{fig:spectrum} PSD of the normalized transmitted (left) and
received (right) signals (100\,GBd, $\bar{\phi}\approx1.1$) obtained
analytically from (\ref{eq:autocrrelation_evolution}) or with numerical
simulations (MB or SpSh with $N=4$): (a) without NLPC; (b) with TX-side
NLPC; and (c) with split NLPC.}
\end{figure}

Fig.\,\ref{fig:spectrum} illustrates the spectral-broadening effect
occurring in the HPOA (Setup~A) for three different NLPC configurations:
(a) without NLPC, (b) with TX-side NLPC, and (c) with NLPC evenly
split between TX and RX. 
For each configuration, the left panel shows the PSD of the transmitted
signal before the HPOA, while the right panel shows the PSD of the
received signal after the HPOA. In these examples, the transmit and
receive filters $H_{\mathrm{TX}}(f)$ and $H_{\mathrm{RX}}(f)$ are
omitted so that the full extent of the spectral broadening can be
observed, while their bandwidths are indicated by black dashed lines
for reference. Numerical simulations based on the split-step Fourier
method (SSFM), averaged over many realizations, are compared with
the analytical prediction based on (\ref{eq:autocrrelation_evolution}).
The results are obtained for probabilistically shaped 64QAM signals
at a launch power of $P=43$\,dBm (corresponding to $\bar{\phi}\approx1.1$).
Probabilistic shaping is implemented either through i.i.d. symbols
with an MB distribution, representative of ideal AWGN-optimal shaping
and practically achievable for large block lengths $N$, or through
SpSh with block length $N=4$, corresponding to the simple LUT implementation
shown in Fig.~\ref{fig:lut45}.

Without NLPC (Fig.~\ref{fig:spectrum}(a)), the PSD of the transmitted
signal is solely determined by the pulse shape $p(t)$, independently
of the adopted shaping scheme, and remains fully contained within
the TX bandwidth. After propagation through the HPOA, however, significant
spectral broadening occurs, causing part of the signal spectrum to
extend beyond the available RX bandwidth and reducing the in-band
PSD. In this condition, RX-side NLPC can mitigate, but not fully compensate,
NLPR, since the out-of-band spectral components are irreversibly lost
after detection. For i.i.d. MB symbols, the spectral broadening is
more pronounced and closely follows the analytical prediction of (\ref{eq:autocrrelation_evolution}).
By contrast, SpSh with $N=4$ exhibits reduced spectral broadening
due to the strong correlation among the amplitudes of the four quadrature
components within each DP symbol, an effect not captured by the analytical
model. This observation suggests that reducing the PAS block length
may help mitigate spectral broadening, although such behavior is not
described by the simplified analytical expression in (\ref{eq:autocrrelation_evolution}),
which assumes i.i.d. symbols.

With TX-side NLPC (Fig.~\ref{fig:spectrum}(b)), spectral broadening
could be virtually removed from the RX signal, but at the expense
of inducing a comparable broadening on the TX signal. Therefore, under
symmetric TX and RX bandwidth constraints, TX-side and RX-side NLPC
are expected to experience similar bandwidth limitations and provide
comparable performance. In this scenario, the impact of spectral broadening
could be reduced by evenly splitting NLPC between TX and RX, as illustrated
in Fig.~\ref{fig:spectrum}(c).

\subsection{System Performance\label{subsec:System-performance-and}}

\begin{figure}
\begin{centering}
\includegraphics[width=1\columnwidth]{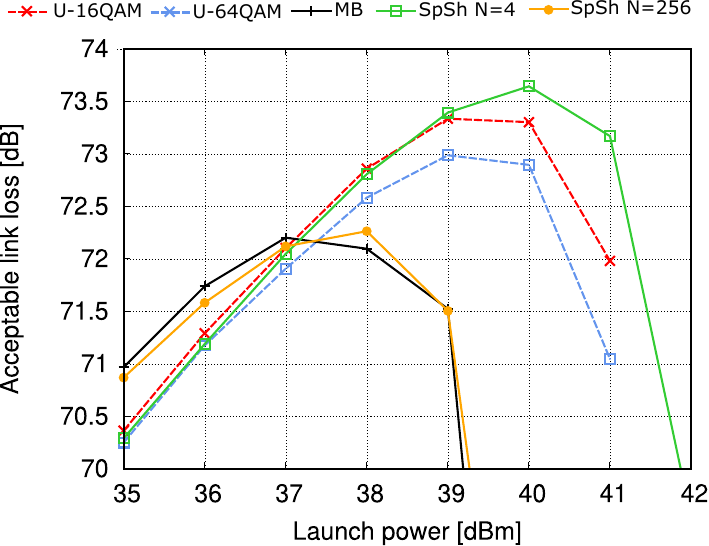}
\par\end{centering}
\caption{\label{fig:PASvsBL}Performance of different uniform and shaped modulations
(target GMI: 3\,bits/2D; NLPC not applied).}
\end{figure}

First, we analyze system performance for different modulation formats
with uniform or shaped distributions. Fig.\,\ref{fig:PASvsBL} shows
the acceptable link loss versus launch power $P$ for a target GMI
of 3\,bits/2D without NLPC, comparing U-16QAM, U-64QAM, i.i.d. symbols
with MB distribution (an ideal benchmark distribution that is nearly
optimal in the linear regime), and SpSh with very short ($N=4$) and
very long ($N=256$) block lengths \cite{gultekin2018Sphereshaping}.
In all cases, PAS is implemented using 4D serial mapping: the $N$
amplitudes generated by the same DM instance are mapped sequentially
onto the quadratures and polarization components of the transmitted
symbols. As a result, each DM instance generates $N/4$ 4D symbols,
meaning that a DM with block length $N=4$ generates a single 4D symbol
\cite{fehenberger2020analysis,civelli2022JLTBPS}. The figure shows
that (i) in the linear regime, ideal MB shaping achieves the best
performance, closely approached by SpSh with $N=256$; (ii) performance
drops in the nonlinear regime due to HPOA-induced nonlinearity; and
(iii) in the nonlinear regime, ideal MB shaping performs worse than
uniform constellations, while SpSh with a very short block length
provides the best performance despite its large rate loss in the linear
regime. This behavior arises from a trade-off between linear shaping
gain and robustness to nonlinear effects for the different formats.
For a given average symbol energy, the linear shaping gain increases
with the constellation entropy, reaching its maximum with MB shaping,
whereas nonlinear effects grow with the energy variations among the
constellation symbols---being absent in a 4D constant-energy constellation,
lower for uniform constellations, and largest for MB. SpSh with $N=256$
closely approximates MB shaping, while SpSh with $N=4$ provides the
best trade-off between linear gain and nonlinear robustness. The
advantage of SpSh with $N=4$ is also evident in Figs.\,\ref{fig:spectrum}(a-c),
which show its reduced spectral broadening compared to MB shaping.

\begin{figure}
\begin{centering}
\includegraphics[width=1\columnwidth]{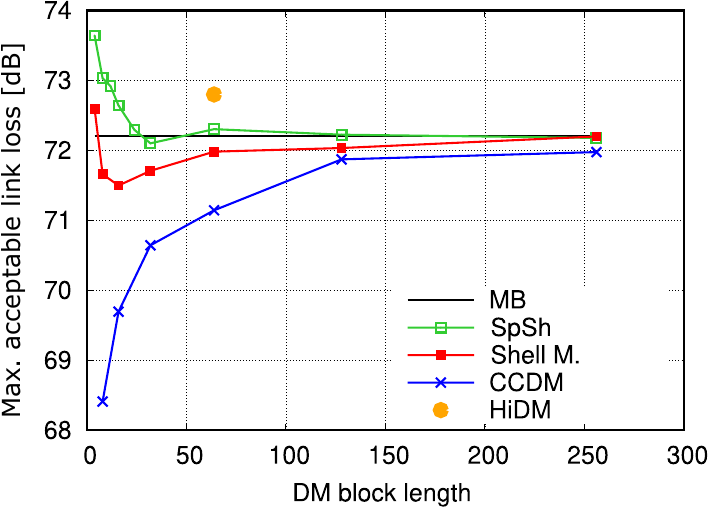}
\par\end{centering}
\caption{\label{fig:PASvsBL2}Performance as a function of the DM block length
for different shaping strategies (target GMI: 3\,bits/2D; NLPC not
applied).}
\end{figure}

A deeper analysis of the behavior of different shaping strategies
is reported in Fig.\,\ref{fig:PASvsBL2}, which shows the maximum
acceptable link loss (at optimal power, taken at $1$dB steps) for
different DM block lengths $N$ for a target GMI of 3~bits/2D and
without NLPC. The figure compares the performance of ideal i.i.d.
symbols with MB distribution, SpSh, shell mapping \cite{civelli2022JLTBPS},
constant composition distribution matcher (CCDM) \cite{schulte2016CCDM},
and one example of hierarchical DM (HiDM) \cite{yoshida2019hierarchicalDM,civelli2020entropy}.\footnote{The HiDM structure is made of $3$ layers, with output amplitudes
$[4,4,4]$, input bits $[1,13,12]$ and alphabet size $[4,64,64]$
and has been optimized to have minimum rate loss under reasonable
constraints and block length.} The figure shows that all techniques converge to the MB performance
for long block lengths, with SpSh outperforming shell mapping, which
in turn outperforms CCDM. This behavior, also observed in conventional
fiber systems, is due to the rate loss of the considered DMs, largest
for CCDM and smallest for SpSh. By contrast, in this scenario, the
optimal block length for SpSh and shell mapping is much shorter ($N=4$)
than in conventional fiber systems \cite{civelli2022JLTBPS,fehenberger2020mitigating,wu2025accounting},
highlighting the importance of correlations between the four quadratures
of a single 4D symbol rather than among adjacent symbols. Conversely,
CCDM achieves its best performance with very long block lengths, due
to the high rate loss incurred at short block lengths. The optimal
SpSh with $N=4$ can be implemented with a LUT with $32$ entries,
sketched in Fig.\,\ref{fig:lut45}, containing the sequences of $N=4$
amplitudes with lowest energy. Finally, HiDM (also implemented with
LUTs) provides a good nonlinear performance, obtained with a longer
block length $N=64$, which may provide advantages in terms of rate
granularity (by steps of $1/N$) and linear performance. Further optimization
of HiDM may provide larger gains. In the following, unless otherwise
stated, SpSh with block length $N=4$ is considered as a shaping scheme
and is simply referred to as NL shaping.

\begin{figure}
\begin{centering}
\includegraphics[width=1\columnwidth]{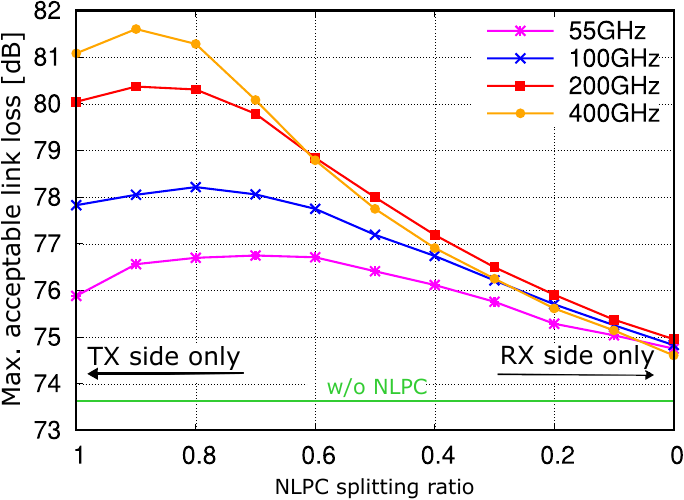}
\par\end{centering}
\caption{\label{fig:LLvsSPLITT}Performance as a function of the NLPC splitting
ratio $\kappa$ for different ADC/DAC bandwidths (oversampling: $n=8$;
target GMI: 3\,bits/2D).}
\end{figure}

Next, we study the performance of the proposed NLPC technique, investigating
its dependence on the splitting ratio $\kappa$ between TX and RX,
on the TX and RX bandwidth, and on the oversampling factor $n$. Fig.\,\ref{fig:LLvsSPLITT}
reports the maximum acceptable link loss as a function of the splitting
ratio $\kappa$, for different bandwidth values $B$, with $n=8$
and a target GMI of 3\,bits/2D. As discussed in Section~\ref{subsec:Nonlinear-phase-compensation},
the effectiveness of NLPC is limited by two factors: bandwidth limitation,
which acts symmetrically at the TX and RX, and signal--noise interaction,
which primarily occurs at the RX. In principle, the first impairment
can be mitigated by increasing the bandwidth and/or by evenly splitting
the NLPC between TX and RX, whereas the second can be reduced by shifting
the NLPC toward the TX. As a result, for a narrow bandwidth of $B=\unit[55]{GHz}$,
the optimal splitting ratio is $\kappa\approx0.6$ (i.e., an almost
even split), yielding a gain of about 3\,dB relative to the baseline
(without NLPC). When the bandwidth increases, only TX-side NLPC improves,
whereas RX-side NLPC remains limited by signal--noise interaction.
Consequently, the optimal splitting ratio shifts toward the TX, resulting
in improved performance.

From a practical standpoint, however, increasing the bandwidth is
generally not a viable option, as it would require allocating additional
spectral resources solely to facilitate NLPC. In fact, such additional
bandwidth could be exploited more efficiently by increasing the symbol
rate, as shown later in this section. Therefore, larger bandwidths
are considered here mainly to provide insight into the operating principles
and fundamental limitations of the technique.

On the other hand, splitting NLPC between TX and RX is a viable implementation
option, albeit at the cost of increased RX complexity. This aspect
is particularly relevant in satellite systems, where onboard processing
resources are typically constrained. Therefore, TX-side NLPC with
limited bandwidth represents the simplest solution. Although suboptimal,
it still provides a significant gain exceeding 2\,dB.

\begin{figure}
\begin{centering}
\includegraphics[width=1\columnwidth]{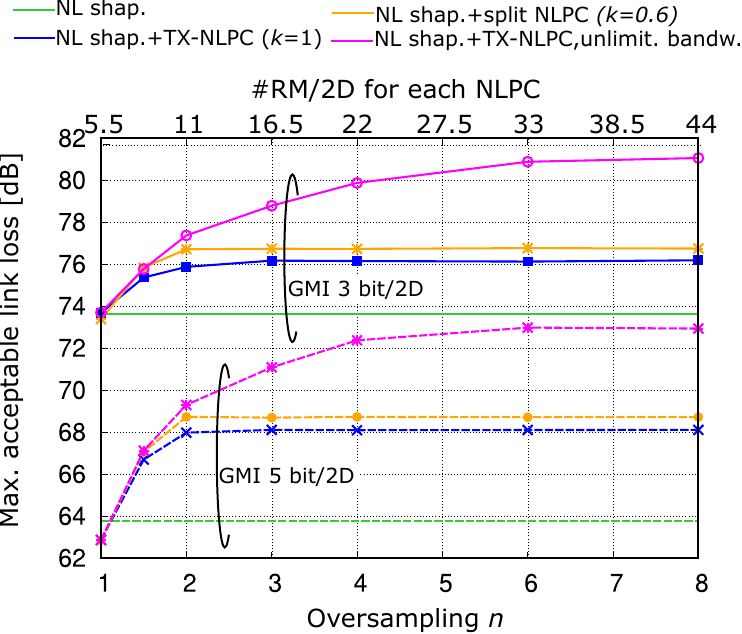}
\par\end{centering}
\caption{\label{fig:LLvsntrx2}Performance as a function of the DSP oversampling
factor $n$ (bottom axis) and corresponding DSP complexity (top axis).
}
\end{figure}

Figure\,\ref{fig:LLvsntrx2} shows the dependence of the performance
on the oversampling factor $n$ (bottom axis) for different NLPC configurations:
TX-side NLPC ($\kappa=1$), split NLPC with optimal splitting ($\kappa=0.6$),
and TX-side NLPC without bandwidth limitations. For each value of
$n$, the corresponding complexity, evaluated from (\ref{eq:complexityNLC}),
is reported on the top axis. The figure shows that the performance
of both TX-side and split NLPC saturates at $n=2$ samples/symbol
when the 55\,GHz bandwidth limitation is included. In this case,
the complexity of each NLPC is low, amounting to only 11\,RM/2D.
As expected, when DAC and ADC bandwidth limitations are removed, the
performance further improves with increasing $n$. A similar behavior
is observed for target GMIs of 3 and 5\,bits/2D.

\begin{figure}
\begin{centering}
\includegraphics[width=1\columnwidth]{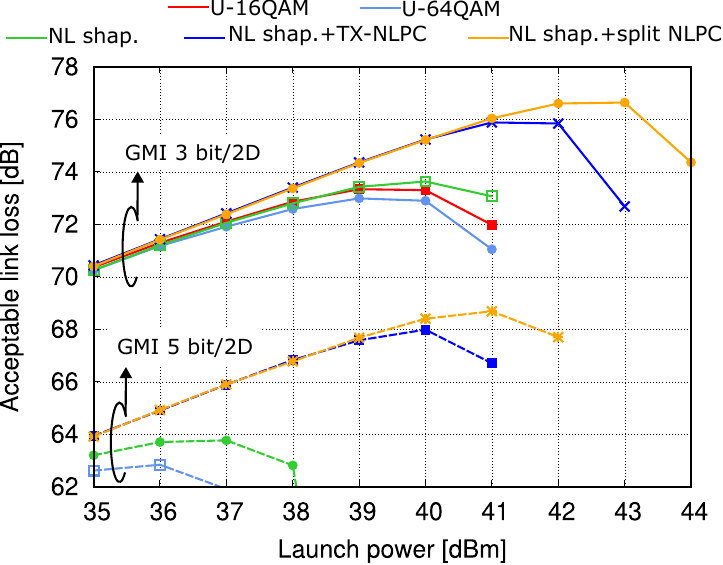}
\par\end{centering}
\caption{\label{fig:LLvsP}Performance of different DSP techniques, for target
GMIs of 5\,bits/2D (dashed) and 3\,bits/2D (solid), with $n=2$.}
\end{figure}

Next, Fig.\,\ref{fig:LLvsP} compares the performance of the different
strategies proposed in this work, considering various shaping and
NLPC configurations: (i) uniform (U) QAM constellations, (ii) constellation
shaping based on PAS optimized for the nonlinear regime with $N=4$
and LUT implementation (NL-shaping), (iii) NL shaping with TX-side
NLPC, and (iv) NL shaping with NLPC split between TX and RX with $\kappa=0.6$,
all under a 55~GHz bandwidth limitation and with $n=2$. As expected,
the acceptable link loss at a GMI of 3\,bits/2D is significantly
higher than at 5\,bits/2D---approximately 6\,dB in the linear regime
and up to 10\,dB in the nonlinear regime---since a lower SNR is
required. Compared to uniform constellations, NL shaping provides
a gain of up to 1\,dB in the nonlinear regime, while yielding approximately
the same performance in the linear regime, as the conventional linear
shaping gain is offset by the large rate loss due to the short block
length. On top of this, NLPC provides additional improvements beyond
NL shaping. In particular, TX-side NLPC yields an extra gain of more
than 2\,dB at a GMI of 3\,bits/2D, and more than 4\,dB at 5\,bits/2D,
relative to NL shaping alone. These gains can be further increased
by about 1\,dB when adopting split NLPC.

So far, assuming DAC and ADC bandwidth limitations of $B=\unit[55]{GHz}$,
we have considered a fixed symbol rate $R=\unit[100]{GBd}$, with
target GMI values of 3\,bits/2D or 5\,bits/2D, corresponding to
overall bit rates $R_{b}=2R\cdot\mathrm{GMI}$ of 600\,Gb/s or 1\,Tb/s,
respectively. However, the same bit rates can also be achieved by
varying the symbol rate $R$ and correspondingly adapting the target
GMI according to $\mathrm{GMI}=R_{b}/(2R)$. Reducing $R$ allows
better containment of spectral broadening within the available bandwidth
$B$, but requires operation at higher GMI and therefore higher SNR.
Conversely, increasing $R$ reduces the required GMI, but leaves less
margin for spectral broadening and may introduce intersymbol interference
even in the linear regime if the signal bandwidth exceeds $B$ (which
occurs at $R\approx\unit[105]{GBd}$).

\begin{figure}
\begin{centering}
\includegraphics[width=1\columnwidth]{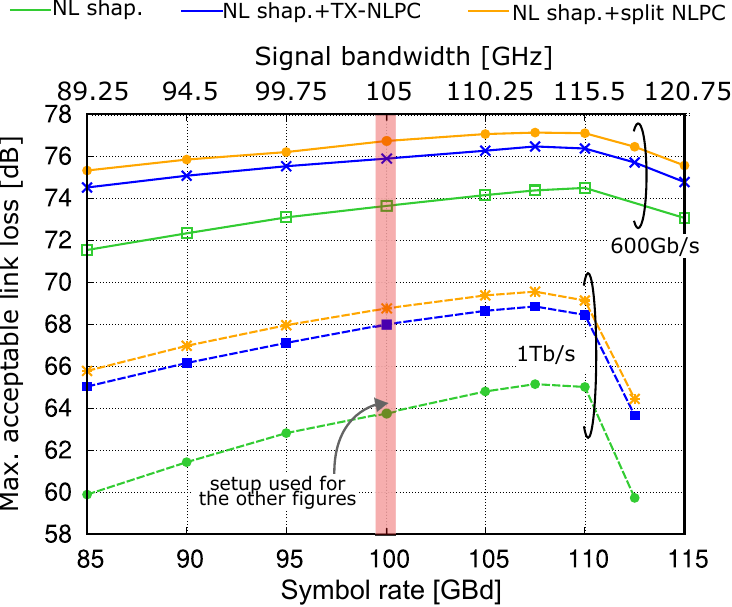}
\par\end{centering}
\caption{\label{fig:LLvsRs}Performance of different DSP techniques as a function
of the symbol-rate with a fixed DAC/ADC bandwidth $B=\unit[55]{GHz}$
and $n=2$. The target GMI changes with the symbol rate to guarantee
a fixed overall bit rate of 1\,Tb/s (dashed) and 600\,Gb/s (solid).
The top axis reports the signal bandwidth.}
\end{figure}

Fig.~\ref{fig:LLvsRs} evaluates this tradeoff by showing the maximum
acceptable link loss as a function of the symbol rate for target data
rates of (i) 600\,Gb/s and (ii) 1\,Tb/s. All results are obtained
with $n=2$, while varying $R$ and selecting the target GMI accordingly.

The figure shows that the largest acceptable link loss is achieved
for symbol rates slightly above 105\,GBd, meaning that reducing the
symbol rate to leave additional margin for spectral broadening is
not advantageous. Instead, it is preferable to fully exploit the available
DAC/ADC bandwidth, and even slightly exceed it, in order to operate
at lower target GMI values and therefore reduce the required SNR.
Although part of the signal spectrum is filtered by the finite DAC/ADC
bandwidth in this regime, the benefit associated with the reduced
GMI requirement more than compensates for the corresponding bandwidth
penalty.

\subsection{Performance Across Amplification Setups and Simplified-Model Validation}

The objective of this section is to compare the system performance
obtained with the different HPOA setups described in Section~\ref{subsec:System-setup}
with that predicted by the simplified dispersionless model in Fig.\,\ref{fig:figSimpleModel},
in order to assess both (i) the validity of the simplified model and
(ii) the effectiveness and generality of the proposed DSP techniques
across different nonlinear propagation scenarios.

\begin{figure}
\includegraphics[width=1\columnwidth]{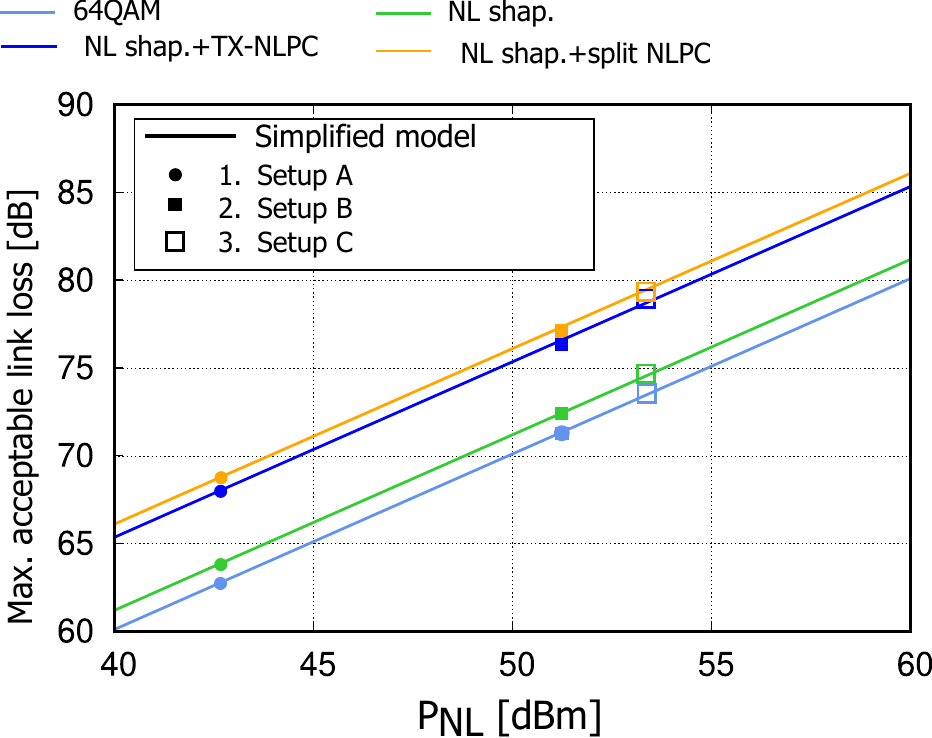}

\caption{\label{fig:gammaeff}Performance as a function of the HPOA characteristic
nonlinear power $P_{\mathrm{NL}}$, for different NLPC techniques
(colors). Markers correspond to specific HPOA configurations, solid
lines are obtained with the simplified model. Target GMI: 5\,bits/2D;
oversampling: $n=2$.}
\end{figure}

Fig.\,\ref{fig:gammaeff} shows the maximum acceptable link loss
versus the characteristic nonlinear power $P_{\text{NL}}=(L_{\text{eff}}\gamma)^{-1}$.
As expected, according to the simplified model (shown with a solid
line), the maximum acceptable link loss increases linearly with $P_{\text{NL}}$,
since nonlinear impairments depend only on the normalized quantity
$P/P_{\text{NL}}$. Therefore, increasing $P_{\text{NL}}$ allows
a higher launch power $P$ and, consequently, a larger acceptable
link loss.

Interestingly, all considered HPOA configurations (markers) follows
the same trend predicted by the dispersionless model, confirming that
their system behaviour can be effectively captured through the single
parameter $P_{\text{NL}}$, despite their different physical implementations.
In this respect, amplifiers characterized by a lower characteristic
nonlinear power $P_{\mathrm{NL}}$ inherently achieve better performance;
however, the proposed DSP techniques remain generally applicable and
provide additional performance gains for each considered setup.

Overall, Fig.\,\ref{fig:gammaeff} confirms the accuracy of the simplified
model in describing HPOA-based uplinks and its suitability as a design
and analysis tool, while also highlighting the robustness of the proposed
DSP strategies across different amplification architectures.

\section{Conclusion\label{sec:conclusion}}

This work has presented a systematic analysis of coherent optical
satellite uplink systems employing high-power optical amplifiers (HPOAs).
To the best of our knowledge, this represents the first comprehensive
investigation of nonlinear effects arising in the HPOA stage within
this emerging application domain. In contrast to long-haul fiber
links, dispersion in the considered scenario is negligible, and nonlinear
distortions are essentially instantaneous, manifesting predominantly
as self-phase modulation and spectral broadening. This fundamentally
alters the interplay between nonlinearity, noise, and bandwidth limitations,
leading to a distinct operating regime compared to conventional fiber
systems.

In the ideal case of unlimited bandwidth, the instantaneous nature
of the nonlinear distortion enables, in principle, perfect nonlinear
compensation. However, in practical systems subject to bandwidth constraints,
nonlinear spectral broadening interacts with filtering and noise,
making compensation more challenging. To address these limitations,
DSP techniques inherited from optical fiber communications have been
adapted and optimized for the satellite uplink context.

To this end, probabilistic shaping and nonlinear phase compensation
(NLPC) strategies have been investigated as complementary approaches
to improve system performance. Probabilistic shaping provides nonlinear
tolerance and rate flexibility. Different shaping strategies have
been analyzed, showing that a simple LUT-based scheme achieves the
largest performance improvement with negligible complexity and minimal
memory requirement, providing up to 1\,dB of gain over uniform constellations.
This is achieved for a very short block length, which reduces rate
granularity. Adjusting the DM block length trades off rate granularity
and nonlinear shaping gain, with HiDM structures offering a promising
solution for further investigation. NLPC enables additional mitigation
of Kerr-induced distortions with low complexity. It can be implemented
at the TX, at the RX, or split between the two and requires only 11
real multiplications per 2D symbol. Overall, the combination of probabilistic
shaping and split NLPC increases the acceptable link loss by about
4\,dB at 600\,Gb/s and 6\,dB at 1\,Tb/s.

The study also reveals that performance gains can be consistently
achieved across a wide range of amplification configurations, despite
their different physical implementations and nonlinear characteristics.
Moreover, these results can be effectively interpreted through a simplified
model that reduces the dependence on the detailed amplifier configuration
to a single equivalent parameter. This provides a compact and accurate
framework for performance prediction, enabling straightforward system
assessment and comparison without requiring explicit modeling of the
underlying amplifier complexity.

Future work will extend the present analysis to wavelength-division
multiplexing (WDM) systems, where inter-channel nonlinear effects
and spectral coexistence will introduce additional design trade-offs.

\appendix{}

We collect the two polarization components of the input field at times
$t+\tau$ and $t$ in the four-dimensional circularly symmetric complex
Gaussian vector 
\begin{align}
\mathbf{z} & =\begin{bmatrix}z_{1}\\
z_{2}\\
z_{3}\\
z_{4}
\end{bmatrix}=\begin{bmatrix}u_{x}(0,t+\tau)\\
u_{y}(0,t+\tau)\\
u_{x}(0,t)\\
u_{y}(0,t)
\end{bmatrix}
\end{align}
with probability distribution
\begin{equation}
p(\mathbf{z})=\frac{1}{\pi^{4}|\mathbf{C}|}\exp\left(-\mathbf{z}^{H}\mathbf{C}^{-1}\mathbf{z}\right)
\end{equation}
where $\mathbf{C}=E\{\mathbf{z}\mathbf{z}^{H}\}$ is the covariance
matrix and $(\cdot)^{H}$ denotes Hermitian conjugation. Assuming
independent and statistically equivalent polarizations, $\mathbf{C}$
has a simple block structure
\begin{equation}
\mathbf{C}=\mathbf{C}_{2}\otimes\mathbf{I}_{2},\qquad\mathbf{C}_{2}=\begin{bmatrix}R_{0} & R_{\tau}\\
R_{\tau}^{*} & R_{0}
\end{bmatrix}
\end{equation}
where $\mathbf{I}_{m}$ is the $m\times m$ identity matrix,
\begin{equation}
R_{0}\triangleq R(0,0),\qquad R_{\tau}\triangleq R(0,\tau)\label{eq:r_0_and_r_tau}
\end{equation}
and $\otimes$ denotes the Kronecker product \cite[Chap. 4]{Horn1991}.
 The determinant and inverse of $\mathbf{C}_{2}$ are
\begin{equation}
|\mathbf{C}_{2}|=R_{0}^{2}-|R_{\tau}|^{2},\qquad\mathbf{C_{2}}^{-1}=\frac{1}{|\mathbf{C}_{2}|}\begin{bmatrix}R_{0} & -R_{\tau}\\
-R_{\tau}^{*} & R_{0}
\end{bmatrix}\label{eq:C2_determinant_inverse}
\end{equation}
Letting 
\begin{equation}
\mathbf{A}=\mathbf{A}_{2}\otimes\mathbf{I}_{2},\qquad\mathbf{A}_{2}=\mathrm{diag}(1,-1)
\end{equation}
the autocorrelation at distance $L$ reads
\begin{align}
R(L,t) & =E\{z_{1}z_{3}^{*}\exp(-j\bar{\phi}\mathbf{z}^{H}\mathbf{A}\mathbf{z})\}\nonumber \\
 & =\frac{1}{\pi^{4}|\mathbf{C}|}\int z_{1}z_{3}^{*}\exp(-\mathbf{z}^{\dagger}\mathbf{D}\mathbf{z})d\mathbf{z}\label{eq:autocorrelation_full_expression}
\end{align}
where we have collected the terms at the exponent in a single quadratic
form with matrix $\mathbf{D}=\mathbf{C}^{-1}+j\bar{\phi}\mathbf{A}$.
Exploiting the block structure of $\mathbf{C}$ and $\mathbf{A}$,
we rewrite $\mathbf{D}=\mathbf{D}_{2}\otimes\mathbf{I}_{2}$, with
\begin{equation}
\mathbf{D}_{2}=\mathbf{C}_{2}^{-1}+j\bar{\phi}\mathbf{A}_{2}=\frac{1}{|\mathbf{C}_{2}|}\begin{bmatrix}R_{0}+j\bar{\phi}|\mathbf{C}_{2}| & -R_{\tau}\\
-R_{\tau}^{*} & R_{0}-j\bar{\phi}|\mathbf{C}_{2}|
\end{bmatrix}
\end{equation}
A direct computation gives
\begin{align}
|\mathbf{D}_{2}| & =\frac{1+\bar{\phi}^{2}|\mathbf{C}_{2}|}{|\mathbf{C}_{2}|}\label{eq:D2_determinant}\\
\mathbf{D}_{2}^{-1} & =\frac{1}{1+\bar{\phi}^{2}|\mathbf{C}_{2}|}\begin{bmatrix}R_{0}-j\bar{\phi}|\mathbf{C}_{2}| & R_{\tau}\\
R_{\tau}^{*} & R_{0}+j\bar{\phi}|\mathbf{C}_{2}|
\end{bmatrix}\label{eq:D2_inverse}
\end{align}
Multiplying and dividing (\ref{eq:autocorrelation_full_expression})
by $|\mathbf{D}|$, we can interpret it as the expectation of $z_{1}z_{3}^{*}$
under a Gaussian distribution with covariance matrix $\mathbf{D}^{-1}$,
divided by the factor $|\mathbf{C}||\mathbf{D}|$. This expectation
simply equals the corresponding term of the covariance matrix $\mathbf{D}^{-1}$,
hence 
\begin{equation}
R(L,t)=\frac{(\mathbf{D}^{-1})_{1,3}}{|\mathbf{C}||\mathbf{D}|}=\frac{(\mathbf{D}_{2}^{-1})_{1,2}}{|\mathbf{C}_{2}|^{2}|\mathbf{D}_{2}|^{2}}\label{eq:autocorrelation_compact}
\end{equation}
where the second equality follows from the properties of the Kronecker
product. Replacing (\ref{eq:C2_determinant_inverse}) and (\ref{eq:D2_determinant})--(\ref{eq:D2_inverse})
in (\ref{eq:autocorrelation_compact}), we eventually obtain 
\begin{equation}
R(L,\tau)=\frac{R_{0}}{\left[1+\bar{\phi}^{2}(R_{0}^{2}-|R_{\tau}|^{2})\right]^{3}}
\end{equation}
Finally, using (\ref{eq:r_0_and_r_tau}) and noting that the normalization
adopted in the Manakov equation (\ref{eq:Manakov}) implies $R_{0}=R(0,0)=1/2$,
we obtain the expression (\ref{eq:autocrrelation_evolution}) reported
in the main text.

The analysis can be extended to the $M$-mode version of (\ref{eq:Manakov})---the
so called \emph{generalized Manakov equation}, which applies for instance
to the propagation in multi-mode fibers in the strong coupling regime
\cite{mecozzi2012nonlinear}. Defining a $2M$-dimensional Gaussian
vector $\mathbf{z}$ with covariance matrix $\mathbf{C}=\mathbf{C}_{2}\otimes\mathbf{I}_{M}$
($M$ independent and statistical equivalent modes), and following
a similar derivation (replacing $\mathbf{I}_{2}$ with $\mathbf{I}_{M}$
everywhere), one obtains
\begin{equation}
R(L,\tau)=\frac{R(0,\tau)}{\left[1+\bar{\phi}^{2}\left((1/M)^{2}-|R(0,\tau)|^{2}\right)\right]^{M+1}}
\end{equation}
For $M=1$, the well-known result for the scalar nonlinear Schr\"odinger
equation is recovered \cite[eq. (4.1.20)]{manassah1991self,agrawal2012fiber}
.

\section*{Acknowledgment}

This work was supported by the European Union - Next Generation EU
under the Italian National Recovery and Resilience Plan (NRRP), Mission
4, Component 2, Investment 1.3, CUP J53C22003120001, CUP B53C22003970001,
partnership on \textquotedblleft Telecommunications of the Future\textquotedblright{}
(PE00000001 - program \textquotedblleft RESTART\textquotedblright ).

\bibliographystyle{ieeetr}

\end{document}